\documentclass[%
 aip,
amsmath,amssymb,
preprint,%
]{revtex4-2}

\usepackage{graphicx}
\usepackage{dcolumn}
\usepackage{bm}
\usepackage{xcolor}
\usepackage[utf8]{inputenc}
\usepackage[T1]{fontenc}
\usepackage{mathptmx}
\usepackage{etoolbox}
\usepackage[nice]{nicefrac}
\makeatletter
\def\@email#1#2{%
 \endgroup
 \patchcmd{\titleblock@produce}
  {\frontmatter@RRAPformat}
  {\frontmatter@RRAPformat{\produce@RRAP{*#1\href{mailto:#2}{#2}}}\frontmatter@RRAPformat}
  {}{}
}%

\makeatother
\begin{document}

\title{Efficient Estimation of Transition Rates as Functions of pH}
\author{Luca Donati}
\email{donati@zib.de}
\affiliation{Zuse Institute Berlin, Takustr. 7, D-14195 Berlin, Germany}
\author{Marcus Weber}
\email{weber@zib.de}
\affiliation{Zuse Institute Berlin, Takustr. 7, D-14195 Berlin, Germany}

\date{\today}

\begin{abstract}
Extracting the kinetic properties of a system whose dynamics depend on the pH of the environment with which it exchanges energy and atoms requires sampling the Grand Canonical Ensemble.
As an alternative, we present a novel strategy that requires simulating only the most recurrent Canonical Ensembles that compose the Grand Canonical Ensemble.
The simulations are used to estimate the Gran Canonical distribution for a specific pH value by reweighting and to construct the transition rate matrix by discretizing the Fokker-Planck equation by Square Root Approximation and robust Perron Cluster Cluster Analysis.
As an application, we have studied the tripeptide Ala-Asp-Ala.
\end{abstract}
\keywords{molecular dynamics, square root approximation, pcca, rate matrix, binding rates, grand canonical ensemble}
\maketitle
\section{Introduction}
Molecular Dynamics (MD) simulations allow us to understand molecular mechanisms at the atomic level and to estimate kinetic properties of molecular systems such as transition rates.
However, with regard to simulations whose objective is to determine physical observables as functions of pH, the strategies developed to date, commonly referred to as constant-pH simulations \cite{Mongan2005, Chen2014, Barroso2017}, require considerable computational resources since at least one MD simulation must be conducted for each pH value of interest.
From a statistical thermodynamic perspective, constant pH simulations sample the Grand Canonical Ensemble, i.e., the collection of Canonical Ensembles of the system opportunely weighted according to the pH of the environment.
By exploiting this property, we have recently developed a novel approach for estimating transition rates as functions of pH, that requires only the sampling of the most probable Canonical Ensembles \cite{Ray2020, Donati2022}.
The simulations are then reweighted in order to construct the Grand Canonical Ensemble distribution for a target pH.
The method makes use of Square Root Approximation (SqRA) to build the transition rate matrix of the system \cite{Bicout1998, Donati2018b, Donati2021, Donati2022b}, and robust Perron Cluster Cluster Analysis (PCCA+) is employed to construct a coarse-grained rate matrix containing the rates between macrostates as functions of pH \cite{Deuflhard2004,  Weber2018}.
Here, we review the theory of the method and present an application to the tripeptide Ala-Asp-Ala, which can be protonated or deprotonated depending on the pH of the surrounding environment.
\section{Theoretical background}
\label{sec:theory}
The theoretical foundations of the method for extracting kinetic information of the Grand Canonical Ensemble from Canonical Ensemble simulations, namely GCEkinCEs, have already been presented in ref.~\cite{Donati2022}. 
Here, we summarize the main concepts essential for understanding and applying the method.
\subsection{The Grand Canonical Ensemble}
Consider the Grand Canonical Ensemble of a molecular system, i.e., the ensemble of possible states $x$ of the state space $\Omega$ of a molecular system that can exchange energy and atoms with the environment at a constant volume $V$, temperature $T$, and chemical potential $\mu$.
The chemical potential represents the energy absorbed or released by the system when the number of atoms changes and depends on the concentration of a species of atoms or particles that favors the exchange of atoms between the system under study and the environment.
For example, in the context of systems that depend on the acidity of the environment, the chemical potential is a function of the pH, i.e. of the concentration of protons, and the associated Grand Canonical partition function is defined as
\begin{eqnarray}
\mathcal{Z}(\mu(\mathrm{pH}),V,T) 
=
\sum_{N=0}^{\infty} e^{\beta N\mu(\mathrm{pH})} Z(N,V,T) 
=
\sum_{N=0}^{\infty} w_N(\mathrm{pH}) Z(N,V,T)
 \, ,
\label{eq:GrandCanEnsemble1}
\end{eqnarray}
where $Z(N,V,T)$ denotes the partition function of the Canonical Ensemble, i.e., the ensemble of \emph{scenarios} of the system, as called in ref.~\cite{Donati2022}, with fixed number of atoms $N$.
The term $e^{\beta N\mu(\mathrm{pH})}$, with $\beta = \nicefrac{1}{kB T}$ and Boltzmann constant $k_B$, is the weight $w_N(\mathrm{pH})$ of a scenario with $N$ atoms at a specific pH value.

In this formulation, the Grand Canonical partition function considers scenarios with any possible combination of atom numbers, from zero to infinity.
Instead, we assume that the system is well defined by a few $S$ representative scenarios, for example, the protonated and deprotonated structures of the system, and approximate the Grand Canonical partition function as
\begin{eqnarray}
\mathcal{Z}(\mu(\mathrm{pH}),V,T) 
\approx
\sum_{n=1}^{S} e^{\beta N\mu_n(\mathrm{pH})} Z(n,V,T) 
=
\sum_{n=1}^{S} w_n(\mathrm{pH}) Z(n,V,T)
 \, ,
\label{eq:GrandCanEnsemble2}
\end{eqnarray}
where the index $n$ denotes a specific scenario.
%
%

The partition function can be used to define the Grand Canonical distribution that describes the probability that a state $x\in \Omega$ occurs at a particular pH value:
\begin{eqnarray}
\pi(x;\mathrm{pH}) 
\approx
\frac{1}{\mathcal{Z}(\mu(\mathrm{pH}),V,T)} 
\sum_{n=1}^{s}
w_n(\mathrm{pH}) \,
\pi_n(x) \, . 
\label{eq:GrandCanDistr}
\end{eqnarray}
%
In eq.~\ref{eq:GrandCanDistr}, the term $\pi_n(x)$ is the Boltzmann weight that a microstate $x$ would have in the Canonical Ensemble of the $n$th scenario, while $w_n(\mathrm{pH})$ is the weight of the scenario due to the pH.
The Grand canonical partition function $\mathcal{Z}(\mu(\mathrm{pH}),V,T)$ acts as a normalization constant.
\subsection{Infinitesimal generators and transfer operators}
The dynamics of the system at a specific pH value can be represented by a time-dependent probability density $\rho(x,t;\mathrm{pH})$ solution of the partial differential equation
\begin{eqnarray}
  \frac{\partial \rho(x,t;\mathrm{pH})}{\partial t}  
  =
  \mathcal{Q}^*(\mathrm{pH}) \rho(x,t;\mathrm{pH}) \, .
  \label{eq:inf1}
\end{eqnarray}
In eq.~\ref{eq:inf1}, the operator $\mathcal{Q}^*(\mathrm{pH})$ is the infinitesimal generator of the propagator $\mathcal{P}(\tau;\mathrm{pH})$ which propagates probability densities by a time lag $\tau$:
\begin{eqnarray}
    \rho(x,t+\tau;\mathrm{pH}) 
    =
    \exp \left(\mathcal{Q}^*(\mathrm{pH})\,\tau \right)\rho(x,t;\mathrm{pH})
    =
    \mathcal{P}(\tau;\mathrm{pH})\rho(x,t;\mathrm{pH})  \, ,
    \label{eq:Propagator}
\end{eqnarray}
with stationary distribution
\begin{eqnarray}
    \lim_{t \rightarrow +\infty} \rho(x,t;\mathrm{pH}) = \pi(x;\mathrm{pH}) \,,
\end{eqnarray}
as defined in eq.~\ref{eq:GrandCanDistr}.
Note that both $\rho$ and $\mathcal{Q}^*$ depend on the pH, as they are defined by the physical properties of the system.
Instead of considering the evolution of probability densities, it is more feasible to consider the evolution of appropriate observable functions $f(x,t;\mathrm{pH})$, for example, indicator functions.
To this end, we introduce the operator $\mathcal{Q}(\mathrm{pH})$, adjoint of the operator $\mathcal{Q}^*(\mathrm{pH})$ with respect to $\pi(\mathrm{pH})$, that defines the partial differential equation
\begin{eqnarray}
  \frac{\partial f(x,t;\mathrm{pH})}{\partial t} =
  \mathcal{Q}(\mathrm{pH}) f(x,t;\mathrm{pH}) \, .
  \label{eq:Q}
\end{eqnarray}
The operator $\mathcal{Q}(\mathrm{pH})$ is the infinitesimal generator of the Koopman operator that propagates functions $f(x,t;\mathrm{pH})$ forward in time:
\begin{eqnarray}
    f(x,t+\tau;\mathrm{pH}) 
    = 
    \exp \left(\mathcal{Q}(\mathrm{pH})\,\tau \right)f(x,t;\mathrm{pH})
    =
    \mathcal{K}(\tau;\mathrm{pH})f(x,t;\mathrm{pH})  \, .
    \label{eq:Koopman}
\end{eqnarray}
The main advantage of the operators $\mathcal{Q}(\mathrm{pH})$ and $\mathcal{K}(\tau;\mathrm{pH})$ is that they allow the determination of physical properties of the system, such as transition rates.
The forms of the infinitesimal generators and transfer operators depend on the underlying equations of motion that drive the dynamics of the system.
Here, we assume that dynamics is well represented by the overdamped Langevin dynamics and eq.~\ref{eq:inf1} is the Fokker-Planck equation.
This kind of dynamics is considered simplistic for a high-dimensional molecular system, but it is often sufficient to represent molecular dynamics in a low-dimensional space, provided the choice of reaction coordinates minimizes the systematic error caused by dimensionality reduction.
Under this assumption, the infinitesimal generator is written as
\begin{eqnarray}
    \mathcal{Q}(\mathrm{pH})
     =
     - \beta \mathbf{D}(\mathrm{pH})
     \nabla F(\mathrm{pH}) \cdot \nabla + \beta^{-1} \Delta \left(\mathbf{D}(\mathrm{pH})\right) \, ,
     \label{eq:Q_ovd}
\end{eqnarray}
where $\mathbf{D}(\mathrm{pH})$ and $F(\mathrm{pH})$ are respectively the diffusion matrix and the free energy surface of the system at constant pH.
The symbols $\nabla$ and $\Delta$ denote, respectively, the nabla and Laplacian operators.
%
%
%
\subsection{Square Root Approximation ad PCCA+}
Consider a discretization of the state space $\Omega$ in $K$ disjoint subsets $\Omega_i$, for example, a Voronoi tessellation whose cell $\Omega_i$ has centered in $x_i$.
The infinitesimal generator $\mathcal{Q}(\mathrm{pH})$ defined in eq.~\ref{eq:Q_ovd} can be discretized into a transition rate matrix $\mathbf{Q}(\mathrm{pH})$ with entries
\begin{eqnarray}
Q_{ij}(\mathrm{pH}) &=& 
\begin{cases}
Q_{ij,\,\mathrm{adj}}(\mathrm{pH})
&\mbox{if  $i\ne j$, and  $\Omega_i$ is adjacent to $\Omega_j$}  \\
0                                           &\mbox{if  $i\ne j$, and  $\Omega_i$ is not adjacent to $\Omega_j$} \\
-\sum_{j=1, j\ne i}^K Q_{ij}(\mathrm{pH})       &\mbox{if } i=j \, .
\end{cases} 
\, ,
\label{eq:RateMatrix}
\end{eqnarray}
where the rates $Q_{ij,\,\mathrm{adj}}(\mathrm{pH})$ between adjacent subsets are estimated by SqRA as
\begin{eqnarray}
Q_{ij,\,\mathrm{adj}}(\mathrm{pH})
&=& 
D_{ij}(\mathrm{pH})\,
\frac{\mathcal{S}_{ij}}{d_{ij}\mathcal{V}_i} \,
\sqrt{\frac{\pi(x_j;\mathrm{pH})}{\pi(x_i;\mathrm{pH})}}  \, .
\label{eq:rate_adjacent}
\end{eqnarray}
In eq.~\ref{eq:rate_adjacent}, the term $\pi$ denotes the Grand canonical distribution defined in eq.~\ref{eq:GrandCanDistr}, $D_{ij}(\mathrm{pH})$ is the diffusion between adjacent subsets $\Omega_i$ and $\Omega_j$, $\mathcal{S}_{ij}$ is the area of the intersecting surface between the subsets, $d_{ij}$ is the distance between the centers of the subsets, and $\mathcal{V}_i$ is the volume of the subset $\Omega_i$.

The rate matrix $\mathbf{Q}(\mathrm{pH})$ contains the infinitesimal rates between subsets of the state space.
However, molecular systems typically exhibit metastability, i.e., it is possible to identify a finite number $n_c$ of macrostates in which the molecular system remains confined for a long period of time, and transitions between them are only rarely observable.
To determine the macrostates, we recommend the use of the PCCA+ method, which, provided the matrix $\mathbf{Q}(\mathrm{pH})$, is able to assign to each subset $\Omega_i$, the probability of belonging to a given macrostate.
These probabilities are organized into membership functions that constitute the matrix $\chi$ of size $K\times n_c$, such that the sum of the rows is equal to 1.
The matrix $\chi$ allows to write the $n_c\times n_c$ rate matrix between macrostates
\begin{eqnarray}
\mathbf{Q}_c(\mathrm{pH})
&=&
(\chi(\mathrm{pH})^\top \mathrm{diag}(\pi(\mathrm{pH})) \chi(\mathrm{pH}))^{-1} \chi(\mathrm{pH})^\top \mathrm{diag}(\pi(\mathrm{pH})) \mathbf{Q}(\mathrm{pH}) \chi(\mathrm{pH})  \, ,
\label{eq:Qc}
\end{eqnarray}
where $\mathrm{diag}(\pi(\mathrm{pH}))$ is a $K\times K$ diagonal matrix, whose diagonal entries are the entries of the vector $\pi(\mathrm{pH})$ approximating the Grand Canonical distribution.
For more details about SqRA and PCCA+, we refer to refs.~\cite{Donati2018b, Donati2021, Donati2022b} and \cite{Deuflhard2004, Weber2018}.
\section{Numerical experiment}
\label{sec:methods}
\subsection{Simulation details}
\begin{figure}[ht]
\begin{minipage}{1\textwidth}
\includegraphics[width=90mm]{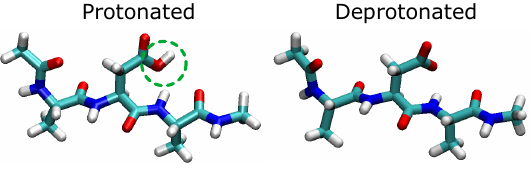} 
\caption{Protonated and deprotonated molecular structure of tripeptide Ala-Asp-Ala.}
\label{fig:1}
\end{minipage}
%
\begin{minipage}{1\textwidth}
\includegraphics[width=45mm]{fig2}
\caption{Occurrence probabilities of the two scenarios defined in eq.~\ref{eq:HHeq}.}
\label{fig:2}
\end{minipage}%
\end{figure}

We studied the tripeptide Ala-Asp-Ala capped with an acetyl group on the N-terminus (ACE) and n-methylamide on the C-terminus (NHMe) to stabilize the molecule.
This peptide exists in two forms, represented in fig.~\ref{fig:1}, that depend on the state of the $\alpha$-carboxylic functional group of aspartic acid: protonated and deprotonated.
The probabilities of occurrence of the two scenarios are estimated from the Henderson-Hasselbalch equations
\begin{eqnarray}
w_p(\mathrm{pH}) &=& \frac{1}{1 + 10^{\mathrm{pH} - \mathrm{pK}_a} } \, , \cr
w_d(\mathrm{pH}) &=& 1 - w_1(\mathrm{pH}) \, ,
\label{eq:HHeq}
\end{eqnarray}
where we used the pK$_a$ value of 3.9 known from experiments \cite{CRCHandbook}.
The functions are illustrated in fig.~\ref{fig:2} and show that the protonated scenario is more likely to occur at pH < 3.9, and vice versa.
In order to estimate the probability distributions $\pi_p$ and $\pi_n$ of the two scenarios and the grand canonical distribution $\pi(\mathrm{pH})$ as function of pH, we carried out MD simulations with the GROMACS 2019.6 package\cite{Spoel2005}, the force field AMBER ff-99SB-ildn \cite{Larsen2010} and the TIP3P water model \cite{Jorgensen1983}.
A velocity rescale thermostat\cite{Bussi2007} was applied to control the temperature and a leap-frog integrator\cite{Hockney1974} was used to integrate the equation of the motion with a timestep of 2 fs.
The length of each simulation was 2 $\mu$s and we printed out the positions every \textsf{nstxout}=500 time steps, corresponding to 1 ps.
Simulations were performed in a NVT ensemble, at temperature of 300 K.

\subsection{Results}
\label{sec:results}
\begin{figure}[ht]
\begin{minipage}{1\textwidth}
\includegraphics[width=70mm]{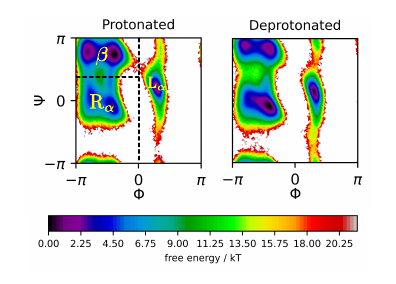}
\caption{Free energy surfaces as defined in eq.~\ref{eq:feps}.}
\label{fig:3}
\end{minipage}
\begin{minipage}{1\textwidth}
\includegraphics[width=67mm]{fig4}
\caption{MSM implied timescales of the protonated and deprotonated scenario.}
\label{fig:4}
\end{minipage}
\end{figure}

As a preliminary analysis of the trajectories generated from the simulations, we estimated the probability distributions $\pi_p(\Phi, \Psi)$ and $\pi_d(\Phi, \Psi)$ of the protonated and deprotonated form of the tripeptide using the dihedral angles $\Phi$ and $\Psi$ of aspartic acid as relevant coordinates.
For this purpose, we built a two-dimensional histogram using a regular space discretization: 36 bins for the $\Phi$ angle and 36 bins for the $\Psi$ angle, for a total of 1296 bins.
Additionally, we estimated the free energy surfaces 
\begin{eqnarray}
    F_p(\Phi, \Psi) = -\frac{1}{\beta}\log \pi_p(\Phi, \Psi)\,, \quad  F_d(\Phi, \Psi) = -\frac{1}{\beta}\log \pi_d(\Phi, \Psi)\, ,
    \label{eq:feps}
\end{eqnarray}
which are illustrated in fig.~\ref{fig:3}.
The two surfaces are similar and exhibit the characteristic regions of a Ramachandran plot: the $\beta$ region, the L$_{\alpha}$ region, and the R$_{\alpha}$ region.
We observe that the $\beta$ and L$_{\alpha}$ regions are connected in the protonated scenario, implying a lower free energy barrier than in the deprotonated scenario.
Furthermore, in the deprotonated form, we note the formation of a barrier between the $\beta$ and R$_{\alpha}$ regions.
In other terms, the torsions around the $\Phi$ and $\Psi$ angles are favored in the protonated form, while are less likely to occur in the deprotonated form.
To confirm this insight, we performed a Markov State Model (MSM) analysis by counting the transitions between the bins of the Ramachandran plot within a lag time chosen in a range between 0 and 1 ns, and building the transition probability matrix $\mathbf{T}(\tau)$ for both scenarios, whose entries $T_{ij}(\tau)$ are the conditional probabilities to observe the system in bin $j$, given it was in bin $i$, after a lag time $\tau$.
From the $\mathbf{T}(\tau)$ eigenvalues, we can estimate the implied timescales, reported in fig.~\ref{fig:4}, that represent the timescales at which the kinetic processes of the system decay.
The graph shows that the MSM implied timescales of both the systems do not depend on the lag time, indicating that the discretization error is negligible and that the two MSMs are a good approximation of the underlying diffusion processes.
The first implied timescale of the protonated state, associated with the transition around the $\Phi$ angle is $t_{1,p}^{\mathrm{MSM}} \approx 4.5 \, \mathrm{ns}$, while the corresponding timescale of the deprotonated state is $t_{1,d}^{\mathrm{MSM}} \approx 25 \, \mathrm{ns}$.
Thus, in the tripeptide's protonated form, the kinetic exchange between the two regions decays five times faster than in its deprotonated state.
We also observe a huge gap between the first and all the other implied timescales, confirming that the dihedral around the left and right halves of the $\Phi-\Psi$ plane is the slowest process of the dynamics, whereas all other processes decay much more quickly.
For more details about MSM theory and application, we refer to \cite{Bowman2014,keller2019}.
%
%

\begin{figure}[ht!]
\includegraphics[width=140mm]{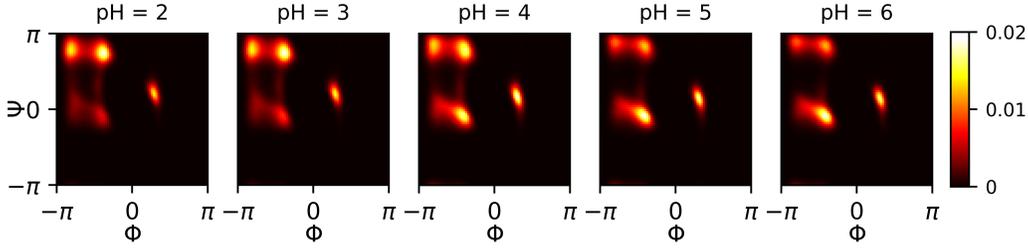}
\caption{Grand Canonical distribution as function of pH.}
\label{fig:5}
\end{figure}

The analysis of free energy surfaces and MSM implied timescales suggests that two metastable states are sufficient to represent the coarse-grained dynamics: one consisting of the $\beta$ and R$_{\alpha}$ region together with $\Phi \in (-\pi, 0]$, and one made up of the L$_{\alpha}$ region with $\Phi \in (0,\pi]$.
However, we decided to assume three metastable states to include the three regions of the Ramachandran plot separately.
First of all, we applied eq.~\ref{eq:GrandCanDistr} with the weights defined in eq.~\ref{eq:HHeq} to build the Grand Canonical distributions for a range of pH values between 2 and 6.
The graphs for five pH values are reported in fig.~\ref{fig:5}, where the dark and bright colors denote regions of the Ramachandran plot with a low and high probability of occurrence at equilibrium, respectively.
At low pH, the $\beta$ and the L$_{\alpha}$ regions are dominant, reflecting the low minima observed in the free energy surface of the protonated form of the tripeptide.
Increasing the pH, the stationary distribution mutates: at pH = 4 the L$_{\alpha}$, the R$_{\alpha}$ and the $\beta$ regions have approximately the same probability; at pH = 6 the $\beta$ region loses its dominance in favor of the L$_{\alpha}$ region.
%

\begin{figure}[ht]
\begin{minipage}{1\textwidth}
\includegraphics[width=50mm]{fig6}
\caption{Diffusion constant along $\lbrace \Phi, \Psi \rbrace$ coordinates as a function of pH.}
\label{fig:6}
\end{minipage}
\begin{minipage}{1\textwidth}
\includegraphics[width=116mm]{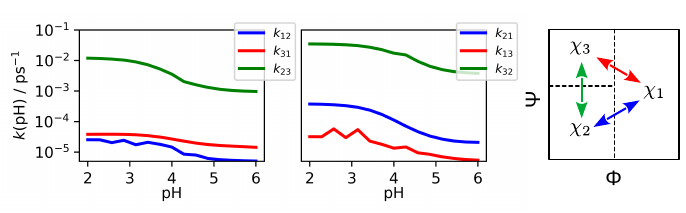}
\caption{Transition rates as functions of pH and schematic representation of the Ramachandran graph divided into three macrostates.}
\label{fig:7}
\end{minipage}
\end{figure}

Afterward, we used the stationary distributions to build the transition rate matrix $\mathbf{Q}(\mathrm{pH})$ applying eqs.~\ref{eq:RateMatrix}, \ref{eq:rate_adjacent}, for each pH value of interest.
As we applied our method to a regular grid, the term $\nicefrac{\mathcal{S}_{ij}}{d_{ij}\mathcal{V}_i}$ simplifies as $\nicefrac{1}{d}$, where $d$ is the size of a bin.
However, as we decided to work in a reduced space represented by the $\Phi$ and $\Psi$ dihedral angles, we do not know the value of the diffusion constant in front of eq.~\ref{eq:rate_adjacent}.  
Thereby, the entries of the SqRA matrix built are not physical rates as their units are expressed in $[\mathrm{rad}^{-2}]$ instead of $[\mathrm{ps}^{-1}]$.
To resolve this issue, we exploited the strict relation between the SqRA rate matrix $\mathbf{Q}$ and the MSM transition probability matrix $\mathbf{T}(\tau)$, and their eigenvalues \cite{Donati2022b}.
Then, we calculated the diffusion constant along $\lbrace \Phi, \Psi \rbrace$ as
\begin{eqnarray}
D = - \frac{1}{\kappa_{1}^{\mathrm{SqRA}} \cdot t_{1}^{\mathrm{MSM}}} \, ,
\label{eq:Diff}
\end{eqnarray}
where $\kappa_{1}^{\mathrm{SqRA}}$ is the second SqRA eigenvalue (the first eigenvalue is always 0 and it is associated with the stationary distribution).
Applying eq.~\ref{eq:Diff} for both the protonated and deprotonated scenario, we obtained the values:
\begin{eqnarray}
D_p &=& 0.064 \, \mathrm{rad}^2 \mathrm{ps}^{-1}
 \cr 
D_d &=& 0.027 \, \mathrm{rad}^2 \mathrm{ps}^{-1}\, ,
\end{eqnarray}
which indicate that a change in pH influences not only the stationary distribution but also the diffusion.
According to Einstein's celebrated work \cite{Einstein1905}, the diffusion constant is related to the variance of the solution of the diffusion equation, then, to determine the relationship between diffusion and pH, we exploited the additive property of variance:
\begin{eqnarray}
D(\mathrm{pH}) = w_p^2(\mathrm{pH}) D_p + w_d^2(\mathrm{pH}) D_d \, ,
\end{eqnarray}
where the weights are given in eqs.~\ref{eq:HHeq}.
The function $D(\mathrm{pH})$ is illustrated in fig.~\ref{fig:6}.

\begin{figure}[ht]
\includegraphics[width=140mm]{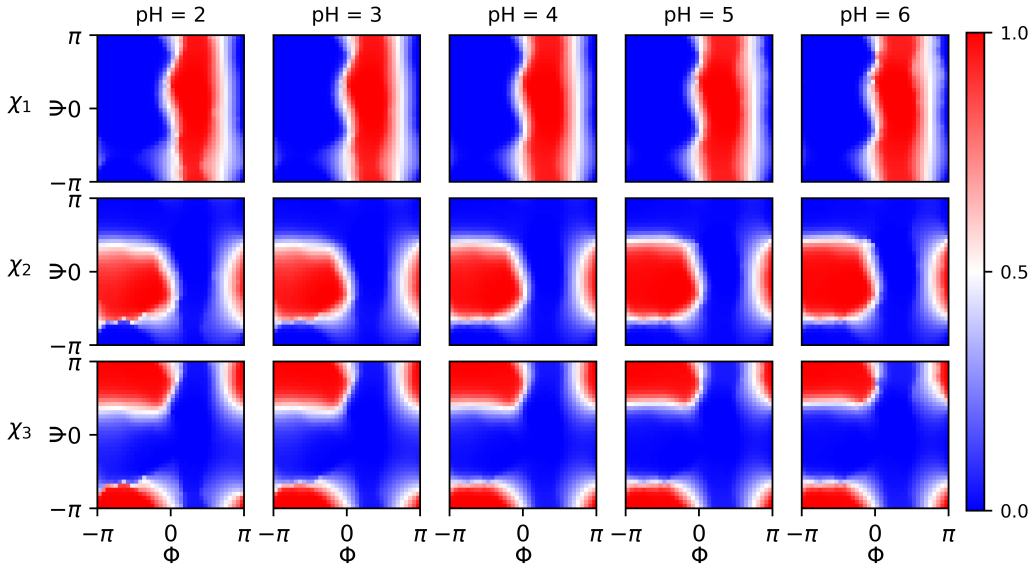}
\caption{Membership functions $\chi$ as functions of pH.}
\label{fig:8}
\end{figure}

The rate matrix $\mathbf{Q}$, multiplied by $D(\mathrm{pH})$, was finally coarse-grained via PCCA+, to build the $3\times 3$ rate matrix of the conformations (eq.~\ref{eq:Qc})
\begin{eqnarray}
\mathbf{Q}_c(\mathrm{pH}) =
\begin{pmatrix}
-k_{12}(\mathrm{pH}) - k_{13}(\mathrm{pH}) & k_{12}(\mathrm{pH}) & k_{13}(\mathrm{pH}) & \\
 k_{21}(\mathrm{pH}) & -k_{21}(\mathrm{pH}) - k_{23}(\mathrm{pH}) &k_{23}(\mathrm{pH})& \\
 k_{31}(\mathrm{pH}) &  k_{32}(\mathrm{pH})  & - k_{31}(\mathrm{pH}) - k_{32}(\mathrm{pH}) & \\
\end{pmatrix} \, ,
\end{eqnarray}
which contains the rates between the Ramachandran regions in both directions.
The rates, as function of the pH, are shown in fig.~\ref{fig:7}.
The highest rates, denoted by $k_{12}$ and $k_{21}$ and represented by a green line in the two graphs, correspond to the transitions $\beta \rightleftharpoons R_{\alpha}$, i.e. the transition within the left half of the Ramachandran plot.
This is the fastest process captured by the coarse-grained model and it occurs with a transition rate of approximately $10^{-2}\,\mathrm{ps^{-1}}$ at low pH, and $10^{-3}\,\mathrm{ps^{-1}}$ at high pH.
The pairs of rates $\lbrace k_{12},\,k_{21}\rbrace$ (blue lines) and $\lbrace k_{13},\,k_{31}\rbrace$ (red lines) denote respectively the slowest transitions  
$\beta \rightleftharpoons L_{\alpha}$ and
$R_{\alpha}\rightleftharpoons L_{\alpha}$.
Since these transitions are rarer, they occur at lower rates, approximately between $10^{-6}\,\mathrm{ps^{-1}}$ and $10^{-3}\,\mathrm{ps^{-1}}$.
Again, we note that the effect of pH is to lower rates, i.e. to raise internal energy barriers and slow down the dynamics.
%
PCCA+ also provides the membership functions $\chi$, i.e. the probabilities that a certain configuration belongs to one of the three macrostates.
The membership functions, plotted in fig.~\ref{fig:8}, are useful to identify the macrostates and the transition states; however, we do not reveal any significant change due to pH.
Thus, we conclude that the effect of pH only influences the transition rates and the probability of occurrence at equilibrium, but does not change the composition of the macrostates.
%
\section{Conclusions}
In this contribution, we presented a method to sample the Grand Canonical Ensemble from a few simulations of the most likely Canonical Ensembles of the system.
The method allows for the estimation of transition rates as continuous functions of the environmental pH, making use of a few simulations.
Furthermore, it can be easily generalized to other environmental variables, such as the salt concentration and the ion concentration which regulate redox systems \cite{Donati2022}.

The method is relevant for the development of new drug design strategies which take into account how the cellular environment influences biochemical processes.
For example, it is indicated for studying ligand-receptor systems whose activation and the emergence of adverse side effects depend on the pH of the cellular membranes hosting the receptor \cite{Spahn2017}.
The only disadvantage is the difficulty in selecting and discretizing reaction coordinates of molecular systems.
For this reason, further research will investigate mesh-free methods such as the recent ISOKANN, which allows estimating membership functions and rates of high-dimensional systems from short MD simulations by means of Neural Network \cite{Rabben2020}.
%

\begin{acknowledgments}
This research has been funded by the Deutsche Forschungsgemeinschaft (DFG, German Research Foundation) Cluster of Excellence MATH+, project AA1-15: ``Math-powered drug-design''.
\end{acknowledgments}
%

%

\end{document}